# Launch Power Optimization in super-(C+L) Systems

Yanchao Jiang[1], Dario Pilori[1], Antonino Nespola[2], Alberto Tanzi[3], Stefano Piciaccia[3],
Mahdi Ranjbar Zefreh[3], Fabrizio Forghieri[3], Pierluigi Poggiolini[1]

1. OptCom, DET, Politecnico di Torino, C.so Duca Abruzzi 24, 10129, Torino, Italy
2. LINKS Foundation, Via Pier Carlo Boggio 61, 10138, Torino, Italy
3. CISCO Photonics Italy srl, via Santa Maria Molgora 48/C, 20871, Vimercate (MB), Italy
*pierluigi.poggiolini@polito.it*

*Abstract*—We investigate launch power optimization in 12-THz super-(C+L) systems, using iterative performance evaluation enabled by NLI closed-form models. We find that, despite the strong ISRS, these systems tolerate well easy-to-implement suboptimal launch power profiles, with marginal throughput loss.

## I. Introduction

Numerous technologies are competing to enhance the throughput of optical links. The extension from C to C+L is already commercially available and is enjoying substantial success. While research is focusing on adding further bands, considerable effort is being directed towards extending the C+L bands into the so-called super-(C+L) ones. In essence, when pushing EDFAs to their limit, amplification can be stretched (in separate units) to the super-L and super-C band, about 6 THz each [1-2]. This represents a significant 20% to 30% increase over the total bandwidth of earlier C+L systems.

One prominent feature of super-(C+L) systems is that, over their 12THz, all fiber parameters become frequency-dependent, including loss, dispersion and non-linearity coefficient. In addition, they are affected by the presence of strong Inter-Channel Raman Scattering (ISRS). These circumstances suggest that a careful optimization of launch power across the spectrum may be essential to avoid substantial performance penalties.

In this paper, we present a detailed study of launch power optimization in SMF super-(C+L) links of lengths 300, 1000 and 3000km. We look at optimized flat launch power, across both bands or per band, and then we consider cubic polynomial laws in each band. We also consider the so-called "3-dB rule" [3-4]. We then provide an accurate estimate of the impact of ISRS on these systems, by turning it off and on, and assessing the throughput change. Finally, we introduce backward Raman amplification, mimicking a commercial 5-pump C+L unit.

To perform the study, iterative system optimization was needed, which in turn required suitable modeling tools. In particular, fast non-linear interference (NLI) closed-form-models (CFMs) are necessary. We used the GN/EGN CFM which was developed in [5-7] and experimentally validated in [8]. It accounts for frequency-dependent loss, dispersion, non-linearity coefficient and ISRS, as well as Raman amplification.

Somewhat surprisingly, our findings indicate that the super-(C+L) systems are quite resilient and robust with respect to the launch power profile used. We also find that, despite ISRS bending individual channel power profiles by several dB's, the actual throughput penalty due to ISRS is quite modest. Finally, we also find that backward Raman amplification is extremely effective in boosting super-(C+L) system performance.

## II. System Assumptions

The simulated test links use parameters derived from an actual 5-span testbed. Each fiber was characterized as for loss, dispersion, non-linearity coefficient and Raman gain profile, across the L and C bands (details in [8]). Each fiber was then analytically stretched to 100km length. The average span loss was about 22.5dB [9]. For the 300 km test-case we used the first three spans of the testbed. For the 1000 and 3000 km links, we replicated the 5-span testbed as many times as needed. The band boundaries are as follows. Super-L: 184.50 to 190.32; super-C: 190.75 to 196.57. EDFAs were assumed with 6dB noise-figure in L and 5dB in C-band. This assumption is somewhat optimistic and factors in some progress in super-(C+L) EDFAs. The WDM signal is 50 channels in each band, with symbol rate 100GBaud, roll-off 0.1 and spacing 118.75 GHz. Modulation was assumed Gaussian-shaped. Throughput was found for each channel by calculating the GSNR (using the CFM) and then resorting to the red curve shown in Fig.1, which is representative of top-performance commercial transponders operated at 100GBaud.

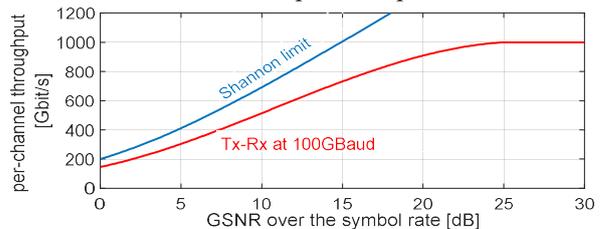

Fig. 1. Per-channel throughput vs. GSNR. Blue: Shannon limit; Red: high-performance Tx-Rx pair at 100 GBaud.

In a first run, the launch power spectrum was described by a cubic polynomial in each band, for a total of 8 free parameters. We optimized them using the overall system throughput as objective function. The results for the 1000km case are shown in Fig.2(a), where the non-linear GSNR (NLI only, red curve), the OSNR (ASE only, green curve) and the overall GSNR (ASE+NLI, blue curve) are shown, together with the optimum launch power spectrum, which we assume to be the same at the start of each span. The markers are individual channels whose performance was calculated on the final configuration using the numerically-integrated EGN-model, as an accuracy check of the CFM result. The plot shows that the optimum launch power is far from constant, growing from 2dBm at the low end of the L-band to 8dBm at the high end of the C-band. GSNR is not flat, decreasing by about 3.5dB, from low-L channels to high-C.

The reason for the GSNR decrease is ISRS, and we see this in Fig.2(b) where we analyzed the exact same system but with ISRS turned off. Both GSNR and launch power are essentially flat, with slight variations due to the frequency-dependence of fiber loss, dispersion and non-linear coefficient. Note also the striking difference in propagation regimes. Without ISRS the system spontaneously adheres to the so-called "3-dB rule" that says that at the optimum launch power the non-linear GSNR is 3 dB higher than OSNR (or equivalently $P_{ASE}$ is 3 dB more than $P_{NLI}$). Instead, when ISRS is present (Fig2(a)), L-band channels are closer to linearity and C-band channels are more non-linear. Surprisingly, despite the quite different curves of the two plots, the throughput penalty due to ISRS is only 0.8 Tb/s, or 1.2%.

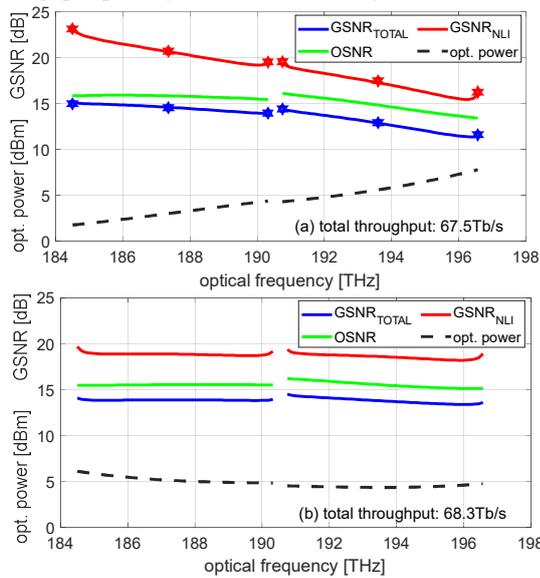

Fig. 2.  1000km SMF max throughput optimization. (a) ISRS on, (b) ISRS off. Markers: results of numerically integrated EGN-model.

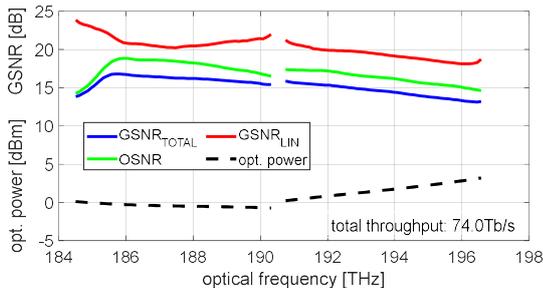

Fig. 3.  3000km SMF max throughput optimization; power and frequency of 5 pump [mW, THz]: 360, 210.6; 320, 208.9; 200, 206.7; 130, 204.5; 180, 200.6.

We then tested three more launch power optimization strategies, with practicality in mind. We tested *flat launch power*, same across bands, or different between the two bands. We also tested *the 3dB rule*, and in this case the optimization target was not max throughput but achieving a 3dB gap between non-linear GSNR and OSNR. Since the optimum polynomial launch power from Fig.2(a) was far from flat, we expected substantial throughput penalty from these non-optimal profiles. Surprisingly, we found very small penalties. The single flat-launch level incurs only a 2% penalty. The dual flat-launch (different L and C-band flat levels) only 1%. Remarkably, the 3dB rule results in just 0.4% less throughput than polynomial. For lack of space, we cannot show the plots of these tests and their replicas at different link lengths, but *all total throughput results* are collected in table 1. Those at 300 km and 3000 km are completely aligned with those at 1000 km, in the sense that small penalties are incurred by the non-optimal profiles.

Finally, we investigated the use of backward Raman amplification. We assumed the pump frequencies and power of a 5-pump telecom unit (caption Fig.3). We then optimized launch power for max throughput (cubic polynomials). The result is shown in Fig. 3 for the 3000km link. The throughput increase vs the same link without Raman is a remarkable +60%. The slight GSNR drop in the low L-band is due to the Raman commercial unit being designed to cover conventional C+L, but the effect is minor and would be easy to correct. The optimum launch power spectrum is remarkably well-behaved.

Table 1.  Total throughput in different optimization strategies

| throughput [Tb/s] | ISRS on | | | | ISRS off |
|---|---|---|---|---|---|
| optimization target | maximum throughput | | | 3dB rule | maximum throughput |
| power spectrum | **cubic** | flat per band | flat both bands | cubic | cubic |
| 300 km | **87.4** | 86.8 | 85.6 | 87.1 | 88.1 |
| 1000 km | **67.5** | 66.7 | 65.3 | 67.1 | 68.3 |
| 3000 km | **46.1** | 45.4 | 44.1 | 45.7 | 46.8 |

### III. Conclusion

Super-(C+L) holds the promise for 12THz of bandwidth. Over such wide band, ISRS is very powerful. This suggests that careful launch power optimization is needed. We investigated this key aspect in depth and found that, while optimization is necessary, even coarse profiles such as constant-per-band are quite effective. The "3-dB rule" is almost optimal. This is a surprising outcome and bodes well for the practical deployment of such wide-band systems. We also found that backward Raman amplification is extremely effective in boosting super-(C+L) system performance (+60% at 3000km), requiring a simple optimized launch power spectrum.

[1]  Y. Frignac et al., "Record 158.4 Tb/s Transmission over 2x60 km Field SMF Using S+C+L 18THz-Bandwidth Lumped Amplification," ECOC 2023, paper M.A.5.2.

[2]  M. van den Hout et al., "Transmission of 138.9 Tb/s over 12 345 km of 125µm Cladding Diameter 4-core Fiber Using Signals Spanning S, C, and L-band," ECOC 2023, paper M.A.5.5.

[3]  G. Bosco et al., "Performance Prediction for WDM PM-QPSK Transmission over Uncompensated Links, " OFC 2011, paper OThO7,

[4]  E. Grellier, A. Bononi, "Quality Parameter for Coherent Transmissions with Gaussian-distributed Nonlinear Noise, " Optics Express, vol. 19, no. 13, pp. 12781–12788, June 20, 2011.

[5]  M. Ranjbar Zefreh, P. Poggiolini, "A Real-Time Closed-Form Model for Nonlinearity Modeling in Ultra-Wide-Band Optical Fiber Links…" arXiv, https://doi.org/10.48550/arXiv.2006.03088

[6]  P. Poggiolini, M. Ranjbar-Zefreh, "Closed Form Expressions of the Nonlinear Interference for UWB Systems," ECOC 2022, p. Tu1D.1.

[7]  Y. Jiang, P. Poggiolini "CFM6, a Closed-Form NLI EGN Model Supporting Multiband Transmission with Arbitrary Raman Amplification," https://doi.org/10.48550/arXiv.2405.08512, May 2024.

[8]  Y. Jiang et al. "Experimental Test of Closed-Form EGN Model over C+L Bands," ECOC 2023, We.C.2.2.

[9]  Y. Jiang et al., "Performance Enhancement of Long-Haul C+L+S Systems by means of CFM-Assisted optimization," OFC 2024, paper M1F.2.